# I. Introduction

Barley (Hordeum vulgare L.), belonging to the Poaceae family, is the fourth major cereal grain cultivated worldwide (Visioni et al., 2019), with a rich profile of carbohydrates, proteins, and dietary minerals like magnesium and selenium. It is abundant in various B vitamins such as thiamine, niacin, and riboflavin. Barley's multifaceted applications extend from human consumption and animal feed to a fundamental role in the production of alcoholic beverages, especially notable in its malted form. This malted barley is crucial in brewing, underscoring its primary role in the malting process, which relies on synchronized and rapid germination (Li et al., 2004). Notably, within the group of monocot plants, barley is often used as model plant for investigating the germination mechanisms. This critical process starts when a quiescent dry seed absorbs water, and achieves on the radicle emergence (Gupta et al., 2019).

The starch composition of barley makes it particularly suitable for producing fermentable sugars, a critical step in alcohol production. Starch is the predominant compound in barley seeds and the source of fermentable sugar production during post-malting step. During barley's germination, the enzymatic hydrolysis facilitates the conversion of this starch into monosaccharides and oligosaccharides, such as glucose, maltose, and maltotriose (Quek et al., 2019). Optimizing barley germination could thus positively impact various sectors, from brewing and distillation to food production and livestock feed. Techniques to enhance germination and seedling growth of barley under normal or stressed conditions, such as hydropriming, have been noted for improving seed germination, particularly under phosphor and zinc deficiencies (Ajouri et al., 2004). Similarly, pre-treatments with abiotic stress factors, like CaCl$_2$ (50 mM), have been shown to activate pre-germination metabolic processes and alleviate salinity stress, thereby benefiting the germination process (Ibrahim, 2016; Kaczmarek et al., 2017).

Cold plasma treatment is a relatively new technique that has gained attention in various industries, including food processing. It is a fast, cost-effective and pollution-free method (Ling et al., 2014). In the context of barley, this involves exposing the grains to ionized gases at low temperature, creating a large array of chemically active species such as reactive oxygen species (ROS) but also UV radiation and electric field (Feizollahi et al., 2020). The benefits of cold plasma on barley include effective reduction or elimination of microbial contaminants (bacteria, yeast, and molds) and reduction of deoxynivalenol (a toxin from Fusarium sp.) without compromising the seeds' quality attributes (Chen et al., 2019; Feizollahi et al., 2020, 2023; Chiappim et al., 2023). Research works also indicate that these effects are obtained without adversely affecting the sensory properties and nutritional composition of the seeds (Misra et al., 2016). Furthermore, plasma treatment has been found to enhance germination and sprouting (Gabdrakhmanov and Tsoi, 2020; Pet'kova et al., 2021), stimulate the production of plant hormones and enzymes, and thus enhance the overall growth of barley plants. Indeed, a 6-min plasma exposure can increase barley seedling weight by 137.5% and boost key metabolites, suggesting its promise as a seed priming technique to enhance plant growth and phytochemical content (Song et al., 2020).

Despite these known benefits, a comprehensive understanding of cold plasma effects on barley's biological, physical, and chemical traits remains elusive. In the context of other cereals like rice, cold plasma can increase the solubility of certain components, such as starch granules (Sun et al., 2022), thereby improving their digestibility and nutrient absorption. It has also been shown to increase the water binding potential and swelling power of corn and tapioca starches (Banura et al., 2018), and modify seed surface properties, which may improve their functionality in various applications, such as in the production of snacks, bread, and other food product production. On the other hand, all these expected benefits are strongly related to the plasma exposure time and may differ for other agriculturally significant seeds, such as peas, as underlined by Kostolani et al. (2021).

While cold plasma treatment holds promise as an innovative approach to cereal pre- and post-harvest processes, more comprehensive research is needed to refine the treatment parameters and evaluate its effects on biological and physical characters of the grains. In this context, our work aims to study the influence of direct and indirect cold plasma treatment on seed reserves (including total sugar, starches, and proteins fractions), seed germination and seedling growth parameters. Our aim is to shed light on the mechanisms underlying cold plasma priming.

## 2. Material and methods

### 2.1. Seed material and plasma treatment

Ardhaoui cultivar barley seeds (Hordeum vulgare L.) were obtained from the Arid Regions Institute of Medenine, Tunisia, with a focus on healthy specimen. Cold plasma (CP) was generated in a dielectric barrier device (DBD) operating in ambient air and supplied by a high voltage generator (16 kV$_{CC}$, 50 Hz). The DBD was composed of two flat alumina electrodes separated by a 1.5 mm thick dielectric barrier (glass). The upper electrode was biased to a high voltage power supply (16 kV$_{CC}$, 50 Hz) while the lower electrode was grounded. The seeds were placed in the 4 mm gap which corresponds to the distance between the lower electrode and the dielectric barrier. Additional information can be found in (Sivachandiran and Khacef, 2017). As sketched in **Fig. 1A**, three types of plasma treatments were achieved on barley seeds: direct plasma treatment of dry seeds (DDS), direct plasma treatment of water-soaked seeds (DWS) and indirect treatment of seeds using plasma-activated water. Regardless of the type of plasma treatment, a 5-min exposure time was consistently applied. Then, the seeds were placed in petri dishes, following protocol detailed in section 2.3.

### 2.2. Diagnostics for plasma phase and plasma-activated water

The electrical properties of the plasma phase were measured using a high-voltage probe (Model P6015A 1000:1, from Tektronix, Beaverton, OR, United States) and a current probe (Model 2877, from Pearson Electronics, Palo Alto, CA, United States) connected to an oscilloscope (Model Wavesurfer 3054 from Teledyne Lecroy company, Chestnut Ridge, NY, United States). The radiative species of the plasma phase were characterized by optical emission spectroscopy (model SR-750-B1-R model from Andor company, Belfast, United Kingdom). The device operated in the Czerny Turner configuration, with a 750 mm focal length while diffraction was achieved with a 1200 grooves·mm$^{-1}$ grating in the visible range. It was equipped with an intensified charge coupled device (ICCD) camera from Andor company (model Istar DH340T). with 2048 × 512 imaging array of 13.5 µm × 13.5 µm pixels).

The plasma-activated water was characterized combining two chemical parameters (pH and electrical conductivity) and two long lifespan reactive species (nitrite and hydrogen peroxide). The pH was monitored using the PS-2147 probe from Pasco Scientific with an accuracy of 0.1. The probe was connected to a computer using a high resolution amplifier and a communication module (PS 3200, Pasco Scientific). Electrical conductivity water was assessed using a conductometer (HI-87314, Hannah) with an accuracy of 1% (full scale). This device was calibrated before measurements using a dedicated standard solution (1413 µS/cm) at 25 °C. Hydrogen peroxide concentration was measured using the titanium oxysufate method. This liquid probe reacts in presence of H$_2$O$_2$ to produce a yellow peroxotitanium complex [Ti(O$_2$)OH(H$_2$O)$_3$]$^+_{aq}$ with an absorbance peak at 409 nm (Reis et al., 1996). Nitrites were quantified by colorimetric assay using Griess reagent.

### 2.3. Water uptake and moisture content of seeds

For both untreated and treated seeds, 2-g samples were weighed and counted. These samples were hydrated with 20 mL of distilled water and incubated at 25 °C for 4 h. After imbibition, seeds were blotted dry and weighed again (Los et al., 2018). Water uptake (WU), expressed in mg/seed, was determined according to formula (1), where FW is fresh weight (mg), DW is dry weight (mg), and n is the seed count per sample. Moisture content (MC) was determined by weighting seeds before and after drying in an oven at 100 °C for 24 h. It is expressed on a % wet basis using formula (2) where W$_i$ and W$_f$ stand for initial and final seed weighs (g) respectively.

$$WU = \frac{FW - DW}{n} \qquad \{1\}$$

$$MC = 100 \times \frac{W_i - W_f}{W_i} \qquad \{2\}$$

## 2.4. Water contact angle measurements

To assess the wettability of barley seed outermost layers, drop shape analysis was performed implementing the sessile drop method. A single droplet (1.5 µL, water) was carefully placed onto the surface of individual barley seed. Then, the water contact angles (WCA) were determined using the ImageJ software (https://imagej.nih.gov/ij/plugins/contact-angle.html).

## 2.5. Germination assays

Germination assays were carried out under four conditions: C (Control), DDS, DWS and IPAW. Fifty seeds were placed on a petri dish with filter paper and moistened with 5 mL of distilled water (C, DDS and DWS) or plasma-activated water (IPAW). Then, germination was monitored during 2 weeks, at 25 °C in darkness, across three replicated experiments. Subsequent growth stages of the seedlings were monitored in a regulated environment within a culture room, maintaining 40% humidity, a consistent 25 °C temperature, and a 16-8 h light-dark cycle.

## 2.6. Germination and seedling growth parameters

Germination was quantified by counting seeds with radicles ≥2 mm. Each treatment, replicated thrice, involved 150 seeds. Germination potential ($GP_\%$), germination rate ($GR_\%$), germination index (GI) and vigor index (VI) were calculated according to formula (3), (4), (5) and (6) respectively (Ling et al., 2014). N is the total number of seeds per petri dish, $n_i$ is number of seeds germinated on one day, $d_i$ is the number of days and L the total length of a seedling in millimeters.

$$GP_\% = \frac{100}{N} \times \left( \begin{array}{c} Number\ of\ seeds \\ germinated\ in\ 3\ days \end{array} \right) \quad \{3\}$$

$$GR_\% = \frac{100}{N} \times \left( \begin{array}{c} Number\ of\ seeds \\ germinated\ in\ 7\ days \end{array} \right) \quad \{4\}$$

$$GI = \sum_i \frac{n_i}{d_i} \quad \{5\}$$

$$VI = GI \times L \quad \{6\}$$

## 2.7. Determination of total soluble sugar and starch contents

The soluble sugars were extracted using 10 mg of fresh matter in 1 mL of 80% ethanol. The content of soluble sugars and starch were determined according to the anthrone method (Staub, 1963).

## 2.8. Determination of storage protein fractions

### 2.8.1. Extraction of storage proteins

To extract storage proteins from barley seeds, a solubility-based fractionation protocol was employed, as described by Elfalleh et al. (2010). A 250 mg milled sample was initially mixed with 5 mL distilled water (pH = 6.5), stirred for 20 min at room temperature, then centrifuged at 10,000 rpm for 15 min. The supernatant, constituting the albumin fraction, was collected as extract 1. The remaining insoluble residue underwent a second extraction using 5 mL of a 5% (w/v) aqueous NaCl solution, producing the globulin fraction as extract 2. Further extractions were performed using a 70% (v/v) aqueous ethanol solution and a 0.2% NaOH solution, resulting in the collection of extract 3 (prolamin fraction) and extract 4 (glutelin fraction), respectively.

### 2.8.2. Proteins storage determination

Protein content in each fraction was determined using Bradford's method the method of (Bradford, 1976). A color reagent was prepared by dissolving 100 mg Coomassie Brilliant Blue G-250 (Sigma-Adrich Co) in 50 mL of 95% ethanol, followed by adding 100 mL of 85% phosphoric acid. The resulting solution was diluted, filtered, and used as the color reagent for protein quantification. BSA standards (Equitech-Bio, Inc., Kerriville, TX), ranging from 0 to 400 µg of protein, were prepared for calibration. Samples and standards were covered with parafilm, mixed, and incubated for 5 min before measuring absorbance at 595 nm using a spectrophotometer.

## 2.9. Measurement of total phenolic contents

Total phenolic content (TPC) of barley ethanolic seeds extracts was determined using the Folin-Ciocalteu reagent method. A calibration standard was performed using gallic acid concentrations (0–600 µg/mL). TPC results were expressed as milligram gallic acid equivalent (GAE) per 100 g of seed fresh weight (mg GAE/100 g FW).

Total flavonoids content (TFC) was measured colorimetrically, based on flavonoid–aluminium complex formation, with maximum absorbance at 430 nm. Each extract (1 mL) was mixed with 1 mL $NaNO_2$ (0.5 M) and 150 µL $AlCl_3$ (0.3 M). After 15 min of incubation at room temperature for 15 min, absorbance was measured at 430 nm. TFC was expressed in mg rutin equivalents per 100 g fresh matter (mg RE/100 g FM), using a rutin calibration curve.

## 2.10. Statistical analysis

Data collected were statistically analyzed using one-way ANOVA with Xlstat software v.2019. Results were presented as means ± standard deviation (±SD). Duncan's test ($p < 0.05$) was performed to evaluate significant differences between treatments. All measurements were performed in triplicate.

# 3. Results and discussion

## 3.1. Characterization of the plasma phase and the plasma-activated water

Whatever the approach considered, the cold plasma of ambient air was generated at a voltage of 8 kV amplitude (i.e. 16 kV$_{dc}$) at a frequency of 50 Hz, according to a sinusoidal profile shown in **Fig. 1B**. The plasma current took the form of a distribution of peaks, sometimes negative, sometimes positive, depending on whether the plasma voltage half-period was negative or positive, respectively. These current peaks can be associated with plasma micro-discharges, whose optical emission is represented as a wavelength spectrum in **Fig. 1C**. This spectrum is dominated by the rovibrational bands of the second positive system of molecular nitrogen (SPS, C$^3\Pi_u$-B$^3\Pi_g$), and also features weakly emissive bands of NO in the γ(A-X) system (220-260 nm region) as well as a characteristic oxygen radical line at 777.4 nm. These reactive oxygen species, among non-radiative others, are likely to improve the germinative properties of seeds (DDS and DWS approaches), as detailed in the following sections.

The same plasma processing can be used to activate distilled water for seed imbibition and irrigation (IPAW approach). Therefore, volumes of 300 mL each were treated by plasma and then characterized through two chemical parameters: pH and electrical conductivity. According to **Fig. 1D**, the pH of distilled water decreases non-linearly from 6.9 to around 5 after 30 min of exposure to cold plasma. At the same time, the electrical conductivity of the liquid increases from 0 to around 1250 μS/cm, indicating the synthesis of new chemical species, notably dissolved ions capable of conducting electricity. These hypotheses were verified by measuring the concentrations of two long-lived chemical species: nitrite and hydrogen peroxide. These two species were considered to be representative markers of the cold plasma effects on distilled water. Naturally, a more exhaustive study could be carried out following the protocols proposed by Judée et al. (2018). The measurements were carried out for plasma exposure times of up to 30 min, consistently using 300 mL volumes. **Fig. 1E** clearly shows the formation of these two species, with values of up to 350 μM and 165 μM of H$_2$O$_2$ and NO$_2^-$ respectively. The greatest increases in their concentrations are obtained during the first 5 min of plasma exposure. Consequently, the plasma effects (according to the DDS, DWS and IPAW approaches) were validated for treatment times of 5 min: a duration which is a good compromise between plasma chemical efficiency on the one hand, and the time constraints linked to the experiments on the other.

## 3.2. Effect of cold plasma on barley seeds germination

Of the three plasma approaches previously mentioned (DDS, DWS, IPAW), we wanted to check whether one of them stood out from the others in terms of its ability to boost germination. Therefore, we monitored the germination potential (GP$_\%$), germination rate (GR$_\%$) and germination index (GI$_\%$) of barley seeds, as shown in **Fig. 2**. These parameters highlighted significant differences among the three plasma groups. First, the DWS treatment totally inhibited germination, since all indicators remained at a zero value. Then, the IPAW treatment revealed to improve the germination rate, with a value of 52%, versus 42% for the control ($p < 0.05$). The third treatment, DDS, was the most effective, leading to a germination rate as high as 78% ($p < 0.05$). Germination potential as germination index followed the same trends as germination rate.

Consistently with these results, the vigor index was significantly increased for the DDS group (543 mm ± 45 mm) as well as the DWS group (553 mm ± 37 mm), compared to the control (131 mm ± 19 mm). Thus, the direct treatment of seeds without water had the strongest stimulatory effect on germination traits., which is consistent with previous studies dealing with tomato (Selcuk et al., 2008), wheat (Jiang et al., 2014), soybean (Ling et al., 2014) or even barley when exposed 15 s to a cold plasma powered in the 40–120 W range (Mazandarani et al., 2020). Regarding barley, the enhancement of seeds germination through DDS and DWS approaches could be explained by several factors:

(i) the reduction of microbial load whether in grains (Los et al., 2018) or in the supplied water,
(ii) (ii) the degradation of the poisonous compound mycotoxin deoxynivalenol produced by some fungi species Feizollahi et al. (2020),
(iii) (iii) changes in the cell membrane's permeability, making it easier for water and essential nutrients to enter the seed, thus facilitating germination,
(iv) (iv) the activation of enzymes within the seed that are crucial for germination.

In contrast, some assumptions could account for the observed germination inhibition resulting from the DWS treatment: (i) First, it is essential to emphasize that the different components of seeds (starch, proteins, lipids), have distinct water absorption capacities. For example, the embryo is rich in proteins and lipids, while the endosperm contains mainly starch which can absorb larger quantities of water. Initially, the seed coat resists water penetration, but its permeability changes upon water absorption, affecting the rate and distribution of water. (ii) The plasma treatment can produce long lifespan ROS (hydrogen peroxide, nitrates, …) as well as short lifespan ROS (hydroxyl radicals, oxygen radicals, …). In fact, ROS could cause a shift in the ion concentration in water. For example, hydroxyl radicals may interact with other ions or compounds in the water, leading to further changes in the electrolytic composition. If the external solution becomes hypertonic, it could hinder the seeds ability to absorb water, i.e. reduce their water uptake. In turn, this osmotic stress could lead to cellular damage as the cells struggle to maintain their internal water balance. (iii) Water-soaked seeds are already in a state of hydration and metabolic activation. Introducing plasma directly could result in an overwhelming amount of RONS being absorbed, causing oxidative stress and damaging the cellular machinery of the seed. (iv) In parallel to the aforementioned assumptions, the plasma treatment may increase seed microbial activity which, in turn, could inhibit germination.

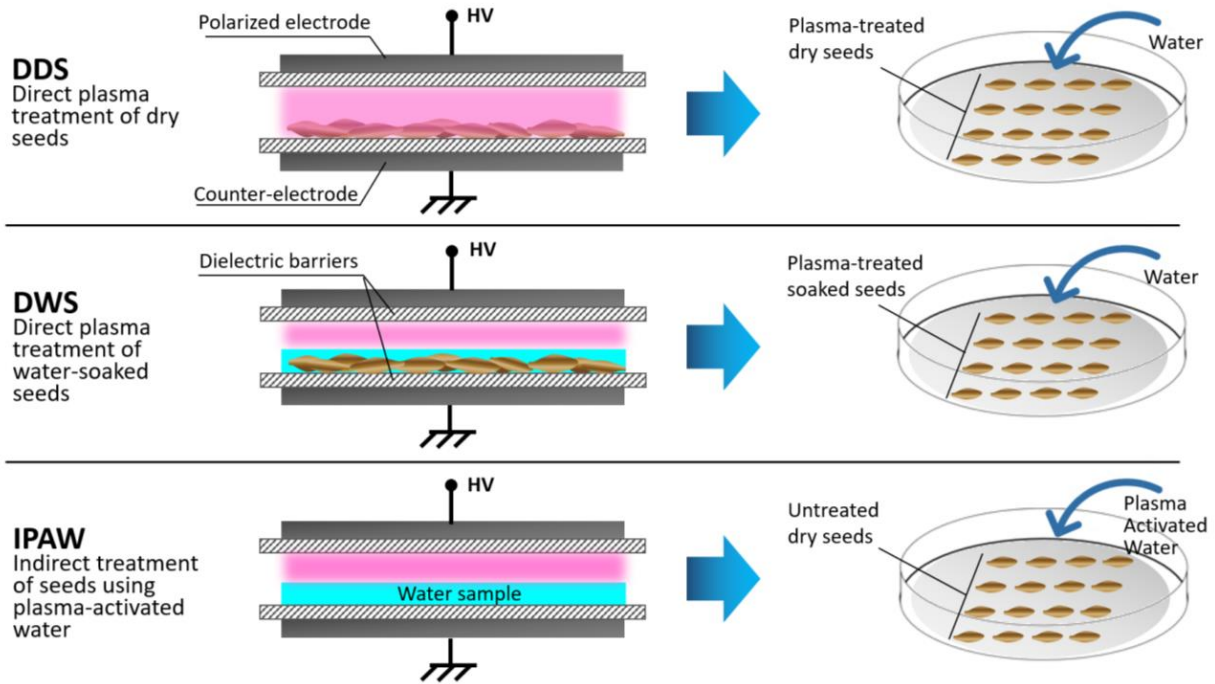

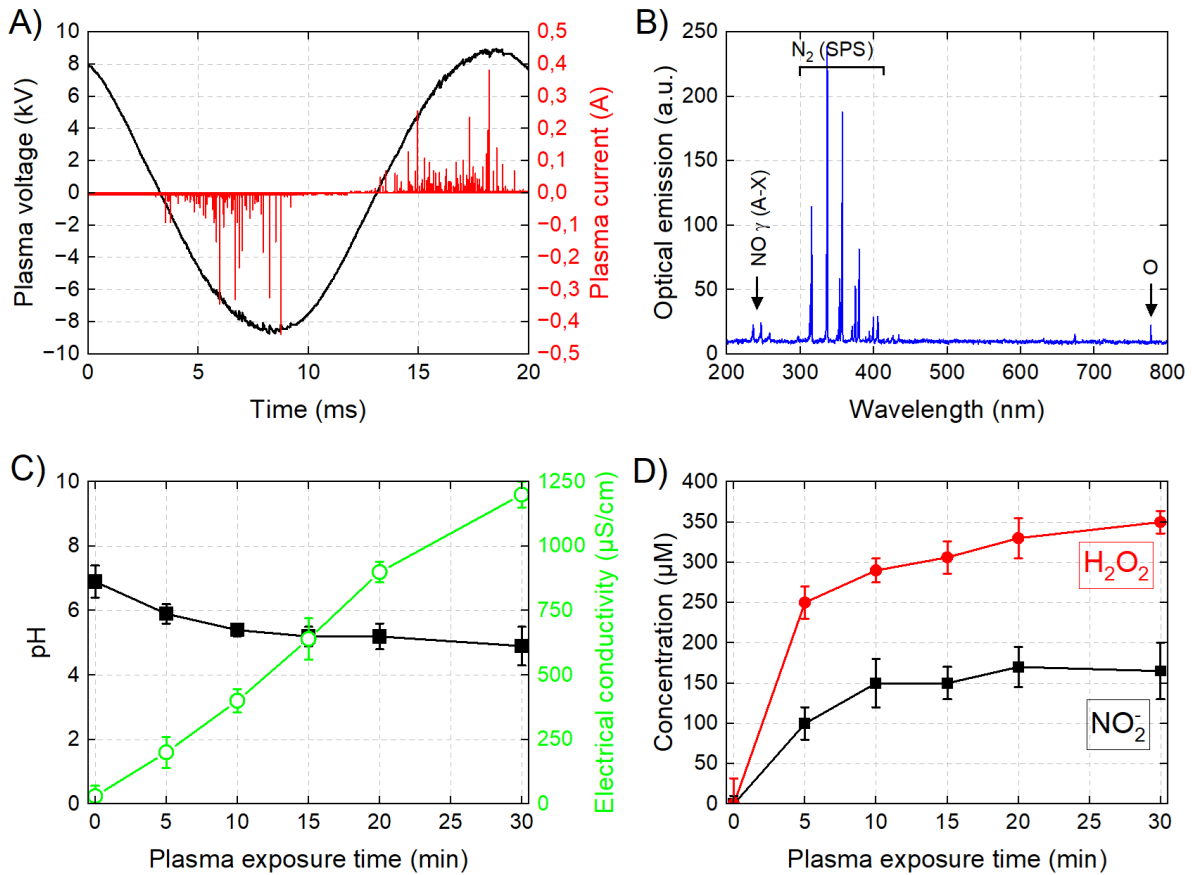

Fig. 1. (A) Overview diagram illustrating the three distinct plasma treatment methods applied to barley seeds, (B) Time profile of voltage and current applied to seeds upon their exposure to air plasma generated in DBD, (C) Optical emission spectrum of the ambient air plasma, (D) pH and electrical conductivity of plasma-activated water as a function of plasma exposure time, (E) Concenration of nitrite and hydrogen peroxide as a function of plasma exposure time.

## 3.3. Variation of water uptake and moisture content

Having established that the three plasma methods influenced germination differently, it became crucial to investigate the causes of these variations. The reactive oxygen species produced by the plasma were delivered to the seeds either before imbibition (DDS) or during it (DWS, IPAW). This made it compelling to examine their impact on the imbibition parameters, especially water uptake (WU) and moisture content (MC), as reported in Table 1.

Regarding water uptake, Control and DDS treatments exhibited analogous profiles with values close to 20 mg/seed. Such similarity implied that the direct plasma treatment of dry seeds (DDS) did not significantly affect the seed's external structures involved in imbibition, at least not in ways that affect water uptake. This result is consistent with the works of Ling et al. (2014) who reported no significant variation of WU for untreated and plasma-treated soybean seeds. However, this trend may depend on seed-type since other results indicate significant increase in the water uptake in the case of wheat seeds after plasma exposure (Los et al., 2018). On the other hand, DWS, where seeds are soaked in water before plasma treatment, displayed significantly diminished water uptake. As reported in **Table 1**, its value is as low as 8.54 ± 2.78 mg/seed versus 20.40 ± 0.28 mg/seed for Control. This reduction may be due to changes to modifications in the seed coat's permeability or a saturation effect from pre-soaking. IPAW, characterized by irrigating seeds with plasma-activated water, fell in between, suggesting a moderate influence on water uptake.

Moisture content presents a distinct trend. In DWS-treated seeds, MC had a value as high as 61.83 %$_{DW}$, which was 7 times more than in untreated seeds. This drastic elevation stood in stark contrast to the other methods and could potentially have both beneficial and detrimental effects. While ensuring hydration, it may also pave the way for potential complications, such as oxygen scarcity within the seed environment. Both Control and IPAW exhibited similar moisture content, affirming that the use of plasma-activated water in IPAW did not significantly alter seed internal hydration. In the case of DDS, moisture content reached a value of 10.26 ± 1.16 %$_{DW}$ (versus 8.70 ± 1.41 %$_{DW}$ for Control). Despite similar water uptake to the control, the slight increase in DDS moisture content suggests possible internal seed changes that retain more water.

By correlating these parameters with germination results, DWS emerged as a potentially harmful treatment. The combination of reduced water uptake and high internal humidity suggested an imbalance that could lead to a decrease of oxygen availability for embryo and the risk of viability loses. This hypothesis was confirmed by the total inhibition of germination observed in seeds treated with DWS. On the other hand, DDS proved to be the most promising, significantly boosting germination rates. The treatment probably induced beneficial changes in the seed, possibly by altering the integument or internal biochemistry, hence improving the propensity to germinate. IPAW was somewhere in between. Although the germination rate was improved compared with the control, it did not reach the efficiency of DDS. Plasma-activated water could introduce beneficial reactive species or induce subtle changes in the dynamics of the interaction between water and seed, thereby promoting germination.

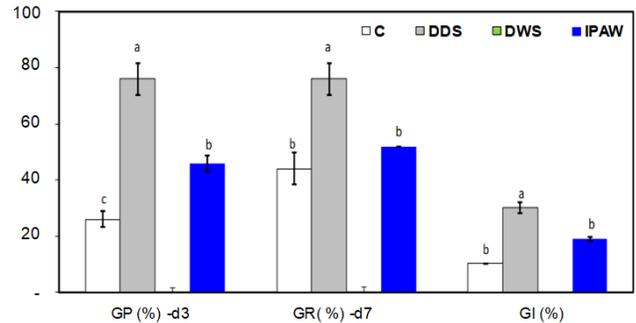

*Fig. 2. Comparative effects of the three plasma methods on barley seed parameters: 3-day germination potential (GP), 7-day germination rate (GR) X-day germination index (GI). Analyzed groups include Control (C), direct plasma treatment of dry seeds (DDS), Direct plasma treatment of water-soaked seeds (DWS) and indirect treatment of seeds using plasma-activated water (IPAW). Error bars represent the standard deviation (SD) from three independent replicates (n = 3). Letters a, b, c indicate significant differences between treatments (p<0.05).*

| Treatment | WU (mg/seed) | MC (%$_{DW}$) | Sugar content (mg/g FM) | Starch content (mg/g FM) |
|---|---|---|---|---|
| C | 20.40 ± 0.28 a | 8.70 ± 1.41 c | 4.83 ± 0.60 a | 499.05 ± 3.18 a |
| DDS | 19.40 ± 0.75 a | 10.26 ± 1.16 b | 5.56 ± 0.25 b | 325.80 ± 10.18 b |
| DWS | 8.54 ± 2.78 c | 61.86 ± 3.53 a | 5.48 ± 0.69 b | 189.00 ± 7.86 c |
| IPAW | 14.15 ± 0.07 b | 8.70 ± 1.41 c* | | |

*Table 1. Effect of plasma treatments on water uptake (WU), moisture content (MC), sugar and starch contents in barley seeds. Mean values in the same row with different letters indicate significant difference between control and plasma treatments (p < 0.05). *; untreated seeds were used with same moisture as CK.*

## 3.4. CP treatment improves hydrophilicity of seeds

Having shown that the three plasma treatments could modify the seeds' bulk water properties, it was relevant to investigate if they also had an impact on the surface water attributes, commonly referred to as wettability properties. Therefore, WCA measurements were achieved in this work to quantify the plasma-induced hydrophilization of the barley seeds. **Fig. 3** clearly shows that all the plasma treatments significantly decreased the WCA values although in different proportions. Indeed, values of 53 °C, 19 °C and 60 °C were obtained for DDS, DWS and IPAW respectively. In the same way, Los et al. (2018) observed a dramatic decrease in the water contact angles on wheat seeds after only 15 s of plasma treatment, hence concluding to a seed surface hydrophilization. The plasma-modification of surface coating was also related to an improvement of seed water permeability, a crucial factor that influences germination parameters (Randeniya and de Groot, 2015). Water absorption, known as imbibition, is a vital stage in ensuring nutrient supply to the growing embryo and generating energy to initiate germination and seedling growth. Plasma treatments improve germination by promoting hydrophilicity at the seed surface and facilitating water uptake, particularly for seeds with thick, less permeable seed coats.

## 3.5. Effect of CP on seed phenolic compounds

Polyphenolic compounds are a distinct class of naturally occurring molecules enriched with multiple phenolic functions. Barley seeds have significant amounts of these molecules that affect the discoloration of food products (Quinde-Axtell and Baik, 2006). **Fig. 4A** shows the effect of cold plasma treatment on the total phenolic content (TPC) and total flavonoids content (TFC) in seeds. While the DDS treatment did not affect the TPC compared with the control, a decline in TFC was observed, likely due to the rapid breakdown of flavonoids. Among different polyphenols, flavonoids were degraded quicker because of their high aptitude to scavenge free radicals generated by plasma (Kumar et al., 2023). For the DWS treatment, there were notable reductions in both TPC and TFC. So far, few studies have addressed the impact of CP on the phenolic compounds. Kumar et al. (2023) reported that the effect of CP on polyphenol compounds essentially depended on the food matrix and plasma process parameters (voltage, frequency, exposure time, feed gas).

composition, we analyzed the soluble sugar (comprising sucrose and β-glucan) and starch contents across different seeds. The findings are reported in **Table 1**. The plasma-treated seeds exhibited increases in sugar content as high as 15.1 % (DDS group) and 13.5% (DWS group) compared to the control. In contrast, the starch content decreased from 499.05 ± 3.18 mg/g FM for untreated seeds to 325.8 ± 10.18 mg/g FM and 189.00 ± 7.86 mg/g FM for DDS and DWS treatments respectively. This indicates that the CP treatment may have promoted the starch breakdown, a crucial step in the germination of barley. From the two forms of starch polymers, Quek et al. (2019) reported that the amylopectin is hydrolyzed more than the amylose during germination. This liberated soluble sugar results from starch degradation under the amylase activity. Indeed, the α-amylase, pivotal for starch hydrolysis, is produced by the aleurone layer of seeds and is induced by the gibberellic acid (GA) released by the embryo. Accordingly, one plausible explanation for the observed changes in starch and sugar contents is the stimulatory effect of CP treatment on the amylase activity. Therefore, the higher germination rate after CP treatment may result from increased availability of sugars released from starch degradation by α-amylase. Studying the amylase activity, the GA changes and the genes families involved at the beginning of germination could elucidate the CP-induced mechanisms.

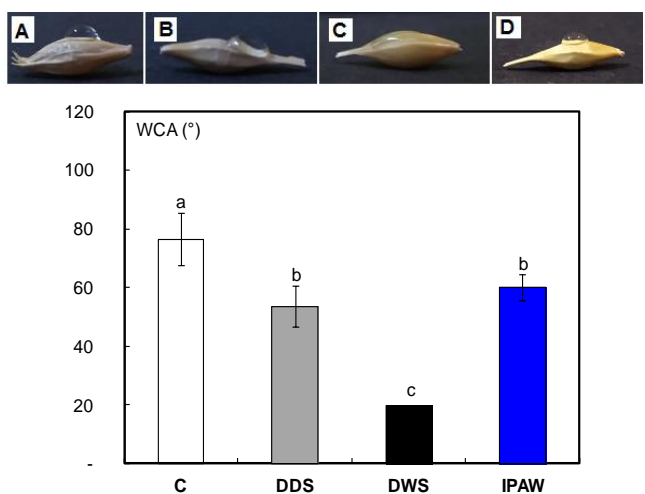

*Fig. 3. Water contact angles (WCA) measurements achieved on four groups of barley seeds: control (C), direct plasma treatment of dry seeds (DDS), direct plasma treatment of water-soaked seeds (DWS), indirect treatment of seeds using plasma-activated water (IPAW). The, b and c letters indicate significant difference between control and plasma treatments (p<0.05). From left to right, photos in inset represent water droplets deposited on barley seeds from C, DDS, DWS and IPAW groups.*

## 3.6. Effect of CP on soluble sugar and starch of barley seeds

In barley seeds, starch constitutes the main storage polysaccharide, accounting for 52%–72% of their dry weight (Lim et al., 2020). These seeds also contain smaller amounts of fructan (1–4%) and mixed-linkage (1,3; 1,4)-β-glucan (10%). To investigate whether the CP treatment affects the seed carbohydrates

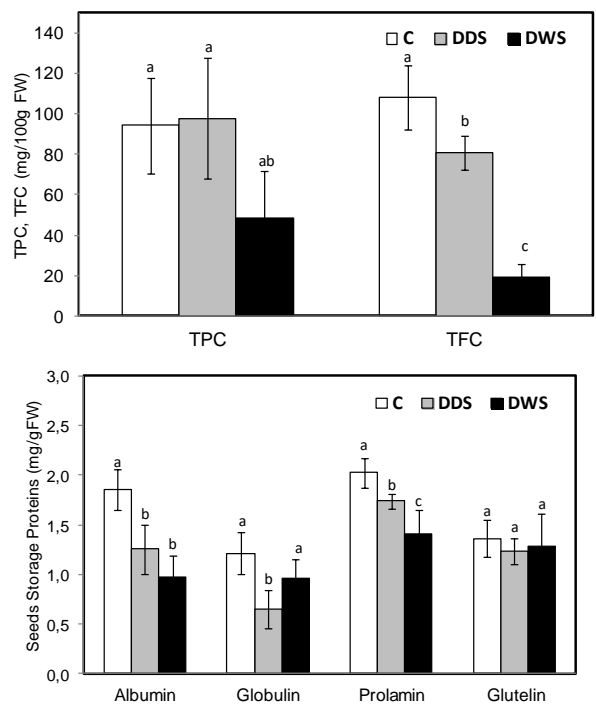

*Fig. 4. Effects of plasma treatment on: (A) the total of polyphenol (TPC) and total flavonoids contents (TFC); (B) Fractions composition of seeds storage proteins of barely seeds: control (C), direct plasma treatment of dry seeds (DDS), direct plasma treatment of water-soaked seeds (DWS). Different letters indicate significant difference between the control, direct treatment and indirect treatment (with water) (p<0.05).*

### 3.7. Cold plasma decreases protein storage

Cereals accumulate reserve proteins in the endosperm tissue of their seeds during maturation. In barley, these proteins are categorized based on solubility into four classes, including globulin (salt-soluble), albumin (water-soluble), prolamin or hordein (alcohol-soluble), and glutelin (not extracted by either solvent). Before germination, these proteins are protected against premature proteolytic attack (Müntz et al., 2001) and maintain stable amounts. During germination and seedling growth, they are degraded by proteases (Kim et al., 2011). **Fig. 4B** displays the variations in storage protein fractions in barley seeds between CP treatment conditions and control (C). For all Osborne protein classes, the highest amounts were found in untreated seeds with values of 1.8 ± 0.1, 1.2 ± 0.21, 2.0 ± 0.15 and 1.36 ± 0.74 mg/g FW for albumin, globulin, prolamin and glutelin respectively. The plasma treatment, applied for 5 min on barley seeds with or without water, decreased albumin, globulin and prolamin fractions. Sun et al. (2020) reported that 5 min of plasma treatment declined the prolamin concentration in wheat seeds to 86.12% of its initial amount. Interestingly, glutelin levels remained consistent across treatments.

From these observations, it can be concluded that the mechanisms protecting against storage protein degradation were probably inhibited by CP treatment, particularly of albumin, globulin and prolamin. It is plausible that the CP treatment triggered partial proteolysis, breaking down these proteins into amino acids and/or subunits. Then, the embryo could use the resulting amino acids as a source of energy. This is confirmed by the increase in germination traits and seedling growth after CP treatment. consequently, a comprehensive proteomic study could be essential in order to uncover the specific impacts of CP on storage proteins degradation.

### 3.8. Seedling growth is enhanced following DDS or DWS procedures

**Fig. 5** shows the impact of cold plasma on the growth of barley seedlings. The shoot length of seedling from seeds treated with DDS and IPAW were 12.0 mm and 18.5 mm, respectively, marking about a 2-fold and 3-fold increase over the control group. Conversely, the DDS treatment did not affect the root growth, as evidenced in **Fig. 5C**. The root lengths of both the control (C) and plasma-treated (DDS) seeds were very close, with measurements of 6.45 mm and 6.1 mm, respectively. However, the root length of seeds irrigated with plasma-activated water (IPAW) was 65.1% higher than the control (p<0.003). Overall, the shoots showed greater increases in length compared to the roots when exposed to the DDS treatment compared with the control.

Similar previous studies have alluded the stimulatory effect of CP on seedlings. For example, Ling et al. (2014) showed that CP treatment had a beneficial impact on plant roots of soybean. Jiafeng et al. (2014) demonstrated that CP could promote the growth of wheat. Also, it has been reported that CP treatment could increase the growth of wheat and oat (Sara et al., 2019). In the present experiment, cold plasma treatments increased the hydrophilicity of barley seeds, due to the grafting of chemical oxygenated functions resulting from the plasma-generated ROS. The WU measurements indicated that this oxidation is not solely limited to the seed surface but is also likely to penetrate into the seeds, resulting in improved nutrients uptake, ultimately resulting in enhanced growth. Likewise, Jiang et al. (2018) suggested that an 80 W CP treatment improved the assimilation of nitrogen and phosphorous in tomatoes by fostering shoot and root growth. In conclusion, whether using DDS or IPAW, the seedling growth of barley could be substantially improved.

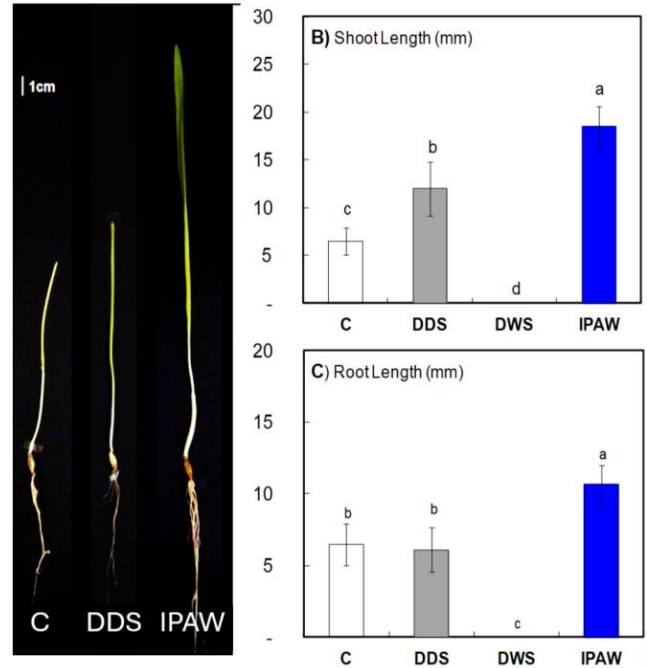

*Fig. 5. Cold plasma affects seedling growth. A: barley seedling after 8 d from sowing. B: shoot length in mm at 8 d, C: Root length in mm at 8 d. Bars represent the standard deviation (SD) of three independent replicates (n = 3). Different letters indicate significant difference between treatments (p<0.05).*

## 4. Conclusion

This comprehensive study of the effects of cold plasma treatments on barley seeds reveals a diverse range of beneficial results. Among the treatment modalities explored, DDS emerged as the most favorable, showing improved germination parameters, probably attributable to beneficial alterations within the seed matrix. Conversely, DWS, characterized by high internal humidity and reduced water uptake, was considered sub-optimal due to potential risks of oxygen deficiency that could compromise seed vitality. In addition, CP treatments significantly increased seed hydrophilicity, an essential determinant influencing the early stages of germination through optimized water uptake. The changes observed in soluble sugar and starch constituents indicate CP's role in stimulating initial germinative activities, such as starch hydrolysis by amylase, thereby increasing the sugar reserves available to the germinating embryo. The observed decrease in protein reserves after CP intervention also implies a potential release of energy for the embryo, in line with the growth-related results. Significantly, the DDS and IPAW treatments stimulated the development of barley seedlings, underlining their potential utility in agronomic efforts. For future studies, a meticulous examination of the molecular and biochemical pathways triggered by CP treatments will be instrumental in refining their deployment for resilient agricultural output.

*Credit authorship contribution statement*
Mohamed Ali Benabderrahim: Writing - review & editing, Writing - original draft, Validation, Investigation, Formal analysis, Conceptualization. Imen Bettaieb: Validation, Formal analysis. Hedia Hannachi: Writing - original draft, Validation, Data curation. Mokhtar Rejili: Validation, Funding acquisition. Thierry Dufour: Writing - review & editing, Writing - original draft, Validation, Supervision, Methodology, Conceptualization.

*Declaration of competing interest*
The authors declare that they have no known competing financial interests or personal relationships that could have appeared to influence the work reported in this paper.

*Data availability*
Data will be made available on request.

*Acknowledgements*
This work was supported and funded by the Deanship of Scientific Research at Imam Mohammad Ibn Saud Islamic University (IMSIU) (grant number IMSIU-RP23006).